# 3D visualization of two-phase flow in the micro-tube by a simple but effective method

X. Fu, P. Zhang[*], H. Hu, C.J. Huang, Y. Huang, R.Z. Wang

Institute of Refrigeration and Cryogenics, Shanghai Jiao Tong University, Shanghai 200240, China

**Abstract:** The present study presents a simple but effective method for 3D visualization of the two-phase flow in the micro-tube. An isosceles right-angle prism combined with a mirror located 45° bevel to the prism is employed to obtain synchronously the front and side views of the flow patterns with a single camera, where the locations of the prism and the micro-tube for clearly imaging should satisfy a fixed relationship which is specified in the present study. The optical design is proven successfully by the tough visualization work at the cryogenic temperature range. The image deformation due to the refraction and geometrical configuration of the test section is quantitatively investigated. It is calculated that the image is enlarged by about 20% in inner diameter compared to the real object, which is validated by the experimental results. Meanwhile, the image deformation by adding a rectangular optical correction box outside the circular tube is comparatively investigated. It is calculated that the image is reduced by about 20% in inner diameter with a rectangular optical correction box compared to the real object. The 3D re-construction process based on the two views is conducted through three steps, which shows that 3D visualization method can be easily applied for two-phase flow research in micro-scale channels and improves the measurement accuracy of some important parameters of two-phase flow such as void fraction, spatial distribution of bubbles, etc..

## 1 Introduction

[*] Corresponding author. Tel.: +86 21 34205505; fax: +86 21 34206814.
E-mail address: zhangp@sjtu.edu.cn (P. Zhang).

Flow pattern visualization is essential for understanding the mechanism of two-phase flow in the micro-scale passages like micro-tubes and micro-channels, etc.. High-speed photography is regarded as one of the most popular visualization techniques in the study of the two-phase flow. Some researchers visualized the two-phase flow in the micro-tube by backlighting arrangement, in which the glass tube was placed between the light source and the camera (Triplett et al. [1], Chen et al. [2]). Others used etching micro-channels in the silicon plate to conduct experiments (Lee et al. [3], Xiong and Chung [4]), in which coaxial lighting was applied in their optical systems by covering a transparent plate. However, the front view of the two-phase flow is the only information obtained in most flow pattern visualization researches in micro-scale passages, so far. The 2D images can only provide partial information and sometimes important information such as the bubble nucleation sites, bubble shapes and spatial distribution of the bubbles, which can provide an in-depth understanding of two-phase flow, are missed or not accurately obtained. Therefore, 3D flow pattern visualization technique, which can help to obtain the accurate spatial details, is indispensable for further quantitative investigation of the two-phase flow.

3D visualization is fulfilled by 3D re-construction which is concerned with the assembly of information obtained in a 2D plane of known orientation into a coherent and comprehensive 3D set of data [5]. The 3D re-construction is widely applied in biomedical imaging and one of the typical examples is CT (computed tomography), by which a large series of two-dimensional X-ray images are taken to generate a 3D image of the inside of an object.

However, in the study of the two-phase flow, there is almost no successful application of 3D visualization in the open literature because of the high expenditure and technical difficulty. Some researchers tried to apply the X-ray tomography in the study of the two-phase flow; however there is great limitation in the temporal and spatial resolution, which can be only applicable for the stationary flow [6]. Schleicher et al. [7] developed a fast optical tomography sensor to study the two-phase flow, which worked similarly to traditional CT. The information obtained by the sensor, was used to re-construct the 2D images (slices). They piled up the slices of a specific cross-section at different time instant and re-constructed the 3D images. Whereas, their 3D images are not virtually the spatial 3D images and could not demonstrate the transient flow characteristics of the whole flow field. Okawa et al. [8] used two synchronized video cameras to visualize the bubble dynamics in a tube of 20 mm in inner diameter. The front view and side view of the two-phase flow were recorded simultaneously. However, the conception of re-construction of the 3D images was not reported in their paper. Takamasa et al. [9, 10] proposed a kind of stereo image-processing method (SIM) to measure the interfacial configuration of bubbly flow under normal and microgravity conditions. They obtained two views from two normal sides of the test section on the same frame by using one video-camera and four mirrors. The method was an efficient approach in enhancing the accuracy and ranges of the measurement of some important parameters in two-phase flow such as void fraction, etc.. Celata et al. [11] modified the optical system of Takamasa et al. [9, 10] by using four mirrors to visualize bubbles, while another four mirrors to lighten the two sides of the rectangular test section. But they did not perform further work to extend the method to other cases with different geometrical configuration such as circular tube, etc.. Also the 3D reconstruction was not conducted though they assumed the bubble shapes as spheroid around the axis to calculate the volume of the bubbles.

The present study provides a simple and effective method to obtain the front view and side view of two-phase flow in the micro-tube synchronously with only a single high-speed video camera. Image deformation due to refraction and geometrical configuration of the test section is quantitatively investigated. With the two views, the 3D re-construction process is demonstrated in detail.

**2 Optical system**

Fig. 1(a) is the typical visualization system of the two-phase flow in a micro-tube at the cryogenic temperature range. The optical system was placed in a high-vacuum cryogenic dewar, in which the illumination and magnification for visualization of the two-phase flow in a micro-tube proved to be a demanding work (Zhang and Fu [12]). Fig. 1(b) shows the enlarged view of 3D visualization system. Back-lighting is employed, in which the transparent glass tube is located between the light source and the camera. An isosceles right-angle prism is arranged with one of the right-angle edges close to the micro-tube. On the other side of the micro-tube, a mirror is placed at an angle of 45° bevel to the prism, which reflects the light from the light source to illuminate the side of the tube. On the focal plane of the high-speed camera, the front and side views of the two-phase flow in the micro-tube are imaged simultaneously. Fig. 1(c) is the bird view of the optical system, in which the image in the prism ($M_1N_1$) is turned 90° clockwise compared to the object points MN, as shown in the visualization field. The image in the prism which is obtained by the camera is actually the side view of two-phase flow in the micro-tube. With carefully adjusting the position of the prism, the front view (AB) and side view (MN) can be clearly obtained synchronously by the camera.

*2.1 Derivation of the relative position of the micro-tube with respect to the prism*

The following paragraphs will derive the exact location of the image with respect to the corresponding object from the perspective of the geometrical optics. Fig. 2 is the optical pathway diagram of the optical system. Incident rays OB and OC are from the object point O. The two incident rays enter into the prism and reflect off the bevel edge of the prism AE. The rays exit at point F and G of the other right-angle edge. Points $O_1$ and O are the corresponding image point and object point which will be recorded by the camera.

The object point O and the image point $O_1$ can be clearly recorded synchronously by the camera only when the two points are imaged simultaneously on the focal plane which indicates that the following relation should be met

$$\overline{GO_1} = \overline{GD} \tag{1}$$

The distance between exit point G and the image point $O_1$ is

$$\overline{GO_1} = (a + nL)/n = a/n + L \tag{2}$$

where $n$ is the refractive index of the prism, $a$ is the length of the right-angle edge of the prism. $L$ is the distance between points O and B. The detailed derivation of the expression is shown in the appendix A.

Eq. (1) comes to the following formula governing the location of the axis of the micro-tube, in which point A is the origin of the coordinates

$$y = a(n-1)/n - x \qquad (3)$$

Fig. 3 shows the relative position according to Eq. (3), in which the refractive index of the prism $n$ is chosen to be 1.5 as an example.

Because the micro-tube is placed on the right side of the prism and the corresponding image points should not exceed the right-angle edge of the prism, it is easy to acquire the definition domain of Eq. (3)

$$x \in [r, a(n-1)/n - r] \qquad (4)$$

where $r$ is the outer radius of the micro-tube.

The range of the length of the right-angle of the prism ($a$) is derived as

$$a(n-1)/n - r \geq r, \quad a \in [2nr/(n-1), +\infty) \qquad (5)$$

The location of the axis of the micro-tube corresponds to the specific line segment $\overline{O'O''}$ in Fig. 3. The corresponding image location $\overline{O_1'O_1''}$ is virtually on the bevel side of the prism, as shown by the red line in the figure. Note that the distance between the axis of the object ($O$) and the corresponding axis of the image ($O'$) is constant ($a(n-1)/n$). It is apparent that the image points $M_1$ and $N_1$ is turned 90° clockwise compared to the object points M and N.

*2.2 Maximal magnification*

For visualization of two-phase flow in micro-scale passages, one of the most important requirements is the magnification of the image. Magnification is the value representing the size of the image captured by the camera compared to the real size of the corresponding object. For a certain prism, the maximal magnification of the image including the two views occurs when the fixed focal plane is full filled by $\overline{OO_1}$ ($a(n-1)/n$) in Fig. 3. And then, the magnification is formulated as

$$M = \frac{f}{\overline{OO_1}} = \frac{nf}{(n-1)a} \qquad (6)$$

where $M$ is the magnification, $f$ is the length of the focal plane of the camera. For prisms with different lengths, the maximal magnification comes when the prism length $a$ has its minimum equal to $2nr/(n-1)$.

2.3 *Experimental results*

Shown in Fig. 4 are the two-phase flow visualization images recorded by the optical system in Fig. 1(a). The test section is a piece of quartz glass tube with the inner diameter of 1.46 mm and is coated with a layer of transparent ITO (Indium Tin Oxide) as the heater to boil the fluid (liquid nitrogen). The parameters of bubble dynamics like the nucleation sites, the bubble departure diameter and the spatial distribution of bubbles, etc., are visually measured at the temporal resolution of 1 ms and spatial resolution of 12 μm (according to the specifications of the present camera), as shown in Fig. 4(a). Fig. 4(b) shows the liquid entrainment observed in the experiment, which results from the interface instability caused by the relative motion between the two continuous phases. The liquid phase, which has greater velocity compared to the vapor velocity, tears the interface and flow through the vapor phase when the inertial force exceeds the surface tension force, as shown by the image at 3 ms in the figure. The spatial location and the phase of the lump could not be specified only based on the front view of the two-phase flow. And the incomplete information may lead to the misinterpretation of the observed phenomenon in the experiments. With the two views, the liquid lump can be verified and the location and shape of the liquid lump can be visually measured.

The two-view optical system has been proven successfully by the demanding cryogenic visualization work and is applicable for room temperature visualization experiments conveniently. The optical system can also be applied for the micro-channel with other non-circular cross-section configurations like rectangular, etc..

## 3 Image deformation due to refraction and geometrical configuration of the test section

Kawahara et al. [13] investigated two-phase flow characteristics in a 100 μm diameter circular tube and found that the inner diameter was almost enlarged by 50% due to optical distortion. However, they did not perform further quantitative study of the enlargement in inner diameter. The image deformation, i.e., the deviation of the image from the corresponding object, is quantitatively investigated in the present study. Shown in Fig. 5 indicates that the obtained images are greatly enlarged in inner diameter compared to the real situation. The inner diameter of the micro-tube obtained from the image is 1.73 mm, which is evaluated by counting the number of pixels that the inner diameter covers in the image, whereas the measured inner diameter is 1.46 mm, indicating that the image is enlarged by about 18.5 % in inner diameter. The image enlargement compared to the real situation lies in two factors: 1) the different refractive indexes of glass wall and the fluid; 2) the geometrical configuration of the test section. Shown in Fig. 6 is the bird view of the optical path concerning the image deformation. The outer diameter of the tube is $r_o$, inner diameter is $r_i$, and the refractive indexes of air outside the tube, the glass and the fluid flowing in the tube are $n_0$, $n_1$ and $n_2$, respectively. Ray AB from object point A penetrates the curved glass wall by twice refraction on the inner and outer surfaces of the glass tube, and then exits the glass tube as ray line CE in parallel direction. $s$ denotes the distance between the center O and the object point $A$; $s'$ denotes the distance between the center O and the image point $A'$.

Apparently, the image point $A'$ is deviated farther away from the center O compared to the object point A. The detailed derivation of the relationship of $s$ and $s'$ is shown in the appendix B. The result is shown in Fig. 7(a), which demonstrates the relationship between the image points and the corresponding object points along the radius. The refractive indexes of glass wall ($n_1$) and liquid nitrogen ($n_2$) are 1.517 and 1.2053, respectively. In the figure, the image is enlarged about 20% in inner diameter compared to the real situation, which agrees well with the experimental results shown in Fig. 5. It is also shown that the deviation of the image point from the corresponding object point can be reduced as the refractive index of the fluid decreases.

The deviation could be neglected when the fluid is gaseous nitrogen ($n_2$=1.000297) as shown in the figure. The effect of the ratio of the wall thickness to the inner radius ($d/r_i$) on the deviation is shown in Fig. 7(b). The data for the three ratios, i.e., 1, 0.1 and 0.01 are plotted in the figure. And it is found that the ratio has no apparent effect on the deviation and can be neglected when the object point is in the range of $s/r_i < 0.7$. However, when the object point approaches to the inner wall ($0.7 \leq s/r_i \leq 1.0$), the effect of the ratio on the deviation is notable.

Kawahara et al. [13] suggested an optical rectangular box to correct the image deformation, in which the box is filled with the same fluid or with some material of the same refractive index as that flowing in the tube. This method is mostly adopted in the conventional cases, in which the deformation due to the curved wall can be almost avoided by this method. However, in the micro-scale cases, installing an optical box filled with some kind of material is complicated and even unacceptable in some cases as the glass tube coated with transparent heating layer. So the optical correction box without filling material, which is more easily realized, is chosen for analysis. The optical path and detailed derivation process for the case of rectangular optical correction box are shown in the appendix B.

Fig. 8(a) shows the results for the cases with the rectangular optical correction box. Inversely, the image is reduced in inner diameter compared to the real object. For the case of $n_2$=1.2053, the image points have about 20% deviation from the corresponding object points. While for the case of $n_2$=1.000297, the image points could be even deviated by 35% in inner diameter. Fig. 8(b) indicates that the size of radius have no effect on the deviation. For radius as small as 0.73 mm and as large as 100 mm, the deviation of image points from the real object points is the same.

Monochromic light is the ideal option for the illumination in the optical system with the prism. If white light is employed, it should be addressed that the chromatic aberration occurs due to the white light decomposed by the prism. Actually, the white light is employed in the present study which may bring about chromatic aberration. For the prism used in the present study, the refractive indexes of red, yellow and blue light are 1.514, 1.517 and 1.522, respectively. The effect of chromatic aberration on the clear imaging is obtained by applying the three refractive indexes into Eq. (3), detailed derivation of which is shown in appendix C. It is seen in Fig. 9 that the chromatic aberration causes 0.59% deviation from the case of yellow light, where the length of the right angle edge of the prism is 12 mm. The calculated deviation distance of 0.048 mm is within the range of depth of field (0.396 mm) in the present optical system. Therefore, the unsharpness of the image caused by chromatic aberration is imperceptible and the white light can be selected as the light source in the present study.

The region, where the front and side views of the flow patterns can be clearly recorded synchronously, can be extended based on Eq. (3) according to the depth of field:

$$y = a(n-1)/n - x \pm k \tag{7}$$

Where $k$ is the depth of field of the lens in the present study. The corresponding two curves are plotted in Fig. 9 in the case of $k$ equal to 0.396 mm. It shows that clear image can be obtained in the region bordered by the two curves of $\pm 9.6\%$.

**4 Re-construction of 3D image**

3D image is re-constructed by taking a series of 2D images (slices). Generally, there are two methods to take slices: one is to get the images of a specific cross-section at different time instant, then re-construct the 3D image by piling up these images; another is to obtain the images of different cross-sections at different time instant, which is mainly applied to micro-CT in biomedical science. However, the former one is less used in the two-phase flow of the micro-tube because the re-constructed image is not actually spatial 3D image and beyond the interest of the study of two-phase flow. The latter method is actually static 3D re-construction and can be only applicable for the stationary flow because of the limitation of spatial and temporal resolution. There is almost no report about the application of the latter method to the transient two-phase flow study in the micro-scale passages. The present study provides a simple but effective method to produce slices from two views depicted in the above section and re-construct the spatial 3D image of the transient two-phase flow in the micro-tube.

The 3D re-construction process is divided into three steps: 1) image segmentation and binarization; 2) taking elliptical slices; 3) piling up slices and re-constructing 3D image.

1) For raw images shown in Fig. 10(a), the background luminance for the front view and side view can be apparently different due to different illumination intensity. Therefore, the image is segmented and different thresholds are applied to the two views in the binary image processing, as shown in Fig. 10(b). Then, a series of image processing algorithms are adopted. Median filter and other filtering algorithms are combined to eliminate the background noises including the stains and scratch on the tube wall. The border of the binary image is extracted. Then the expansion matching method based on the morphology is applied to close the border. Afterwards, the existing holes in the contour are filled. There are still other isolated lines or points connected to the border. Border clearing algorithm is used to clear the border and the final result of step 1 is shown in Fig. 10(c). It is seen that vapor regions such as small bubbles, elongated bubbles are specified by white color, which are ready for taking slices in the next step. It should be pointed out that some uncertainty may be introduced in step 1. The uncertainty is mainly caused by different thresholds for the image binarization, which may cause a maximal uncertainty of 2.3% in major and minor axes.

2) Because the shape of the bubble of two-phase flow is mainly determined by the surface tension force, which acts to minimize the interfacial area and tends to keep bubbles retaining its spheroid shapes. Therefore, it is reasonable to suppose that the circumference of the horizontal cross-section of the bubble shapes elliptically. For obtaining the bubble slices, each row of the binary image is scanned from left to right at the spatial resolution of one pixel, during which the lengths of the continuous white pixels are calculated. For the case in the Fig. 10(c), there are two segments of continuous white pixels for each row of scan, by which the major axis and minor axis can be determined along with the origin of the ellipse. One slice of the bubble is then accomplished by drawing elliptical contour, as shown in Fig. 11(a). And the process is repeated for each row from top to bottom until the whole image is scanned. There are 716 scans totally, in which 457 elliptical slice planes obtained. The spatial distribution of some slices is shown in Fig. 11(b). The data typically containing all the slice planes is saved for further process.

3) The 3D image can be obtained by piling up these slices around a specified axis using the patch algorithm. The 3D image can be displayed in any orientation. Smooth lighting is added and the tube wall is incorporated to locate the two-phase flow, as shown in Fig. 12(a).

The image deformation correction demonstrated in Section 3 can be incorporated in the 3D re-construction process. The elliptical slices taken in step 2 are corrected according to the deviation calculation in Section 3. So the modified slices of the real object are obtained for 3D re-construction in step 3. Fig. 12(b) shows the corrected 3D real object. There are other methods to correct the image deformation such as adding rectangular optical box to the test section, designing special lens for image recording system, etc.. Compared to these methods, the 3D re-construction method shown here is simpler and does not need complicated modification to the test section. Moreover, the re-constructed 3D image can demonstrate more detailed information in three dimensional space, which could be valuable in improving the measurement accuracy of some parameters such as void fraction, spatial distribution of bubbles, etc..

The advantage and limitation of the above 3D re-construction method should be addressed. Generally, the surface tension force is dominant compared to other forces in the two-phase flow in micro-scale passages. So ellipse assumption of the bubble shape and stacking enough slices can be used to approximate the bubble and 3D re-construction method can have good accuracy and benefits to quantitative investigation of bubble dynamics in micro-tubes. The method can also be extended to the two-phase flow system in which the shape of the dispersed phase can be predicted. However, the above method is limited in some unusual cases of two-phase flow system, in which the dispersed phase such as bubbles or droplets have irregular and unpredictable shapes.

## 5 Summary

The present study presents a simple but effective method for 3D visualization of the two-phase flow in the micro-tube. Installing an isosceles right-angle prism and a mirror on the two sides of the micro-tube, the front view and side view of the two-phase flow in the micro-tube are obtained synchronously by a single high-speed video camera. The relative position of the micro-tube with respect to the prism is specified, where the axis of the image is on the bevel edge of the prism and the distance between the axis of the object and the corresponding axis of the image is constant ($a(n-1)/n$). The optical design is proven successfully by the visualization study of flow boiling of liquid nitrogen in a micro-tube.

The image deformation due to refraction and geometrical configuration of the test section is quantitatively investigated. The image is enlarged by about 20% in inner diameter compared to the real situation for the case of liquid nitrogen in the present study. As the refractive index of the fluid increases, the image is enlarged more greatly for the case of circular tube. The image deformation by adding a rectangular optical correction box outside the circular tube is comparatively investigated. It is calculated that the image is reduced by about 20% in inner diameter with a rectangular optical correction box compared to the real situation. Moreover, the image is reduced more greatly in inner diameter as the refractive index of the fluid decreases. The size of the inner radius is found to have no apparent effect on the image deformation.

The 3D re-construction process is divided into three steps. Firstly, the image is segmented into two regions and then binarized. Afterwards, slices are taken by scanning the whole image based on the ellipse assumption of the bubble shapes. Finally, the 3D image is obtained by piling up slices. The deformation can be corrected in the 3D re-construction, which can improve the measurement accuracy of some important parameters of two-phase flow such as void fraction, spatial distribution of bubble, etc..

**ACKNOWLEDGMENTS**

This research is supported by National Natural Science of Foundation of China under contract No. 50776057 and NSFC-JSPS co-operative project under contract No. 50911140104.

**Reference**

[1] Triplett K A, Ghiaasiaan S M, Abdel-Khalik S I, Sadowski D L 1999 Gas–liquid two-phase flow in microchannels Part I: two-phase flow patterns *Int. J. Multiphase Flow* 25 377-94

[2] Chen L, Tian Y S, Karayiannis T G 2006 The effect of tube diameter on vertical two-phase flow regimes in small tubes *Int. J. Heat Mass Transfer* 49 4220-30

[3] Lee M, Wang Y Y, Wong M and Zohar Y 2003 Size and shape effects on two-phase flow patterns in microchannel forced convection boiling *J. Micromech. Microeng.* 13 155-64

[4] Xiong R Q and Chung J N 2007 An experimental study of the size effect on adiabatic gas-liquid two-phase flow patterns and void fraction in microchannels *Physics of Fluids* 19 033301

[5] Baldock R and Graham J 2000 Image Processing and Analysis: A Practical Approach Oxford University Press (UK).

[6] Reinecke N, Petritsch G, Schmitz D and Mewes D 1998 Tomographic Measurement Techniques -Visualization of Multiphase Flows *Chem. Eng. Technol.* 21 7-18

[7] Schleicher E, Silva M J da, Thiele S, Li A, Wollrab E and Hampel U 2008 Design of an optical tomograph for the investigation of single and two-phase pipe flows *Meas. Sci. Technol.* 19 094006

[8] Okawa T, Ishida T, Kataoka I, Mori M 2005 Bubble rise characteristics after the departure from a

nucleation site in vertical upflow boiling of subcooled water *Nuclear Engineering and Design* 235 1149-61

[9] Takamasa T, Kondo K, Kawase M, Rezkallah K S, Clarke N K 1997 Measurement of interfacial configuration of bubbly flow under normal and microgravity conditions using stereo image-processing method *Transactions of the Japan Society of Mechanical Engineers, Part B* 63 396-403.

[10] Takamasa T, Tomiyama A 1999 Three-dimensional gas-liquid two-phase flow in a C-shaped tube *Ninth Inernatinal Topical Meeting on Nuclear Reactor thermal Hydraulics,* San Francisco, California, October 1999: 3-8

[11] Celata G P, Cumo M, D'Annibale F, Marco P D, Tomiyama A, Zovini C 2006 Effect of gas injection mode and purity of liquid on bubble rising in two-component systems *Experimental Thermal and Fluid Science* 31 37-53.

[12] Zhang P, Fu X 2008 Two-phase flow characteristics of liquid nitrogen in vertically upward 0.5 and 1.0 mm micro-tubes: Visualization studies Cryogenics doi: 10.1016 /j. cryogenics. 2008.10.017.

[13] Kawahara A, Chung P M Y, Kawaji M 2002 Investigation of two-phase flow pattern, void fraction and pressure drop in a microchannel *Int. J. Heat Mass Transfer* 28 1411-35

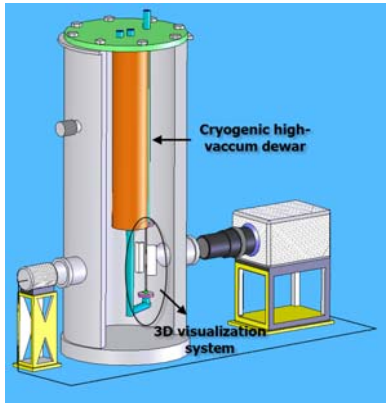 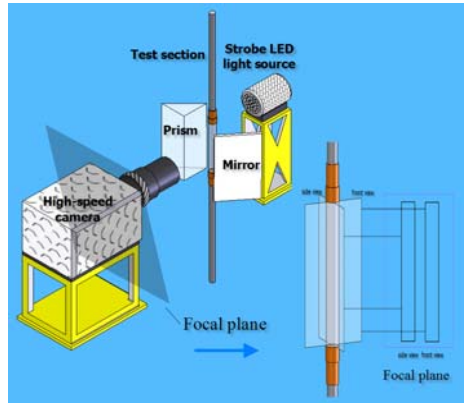 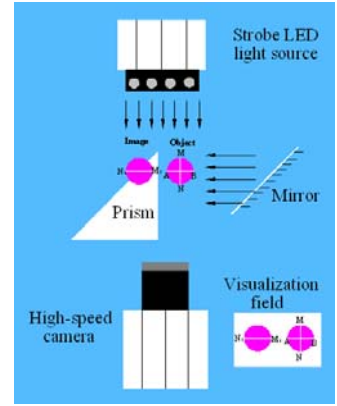

(a) visualization system in cryogenic application   (b) 3D visualization system   (c) Bird view of 3D visualization system

Fig. 1 The arrangement of the 3D optical system

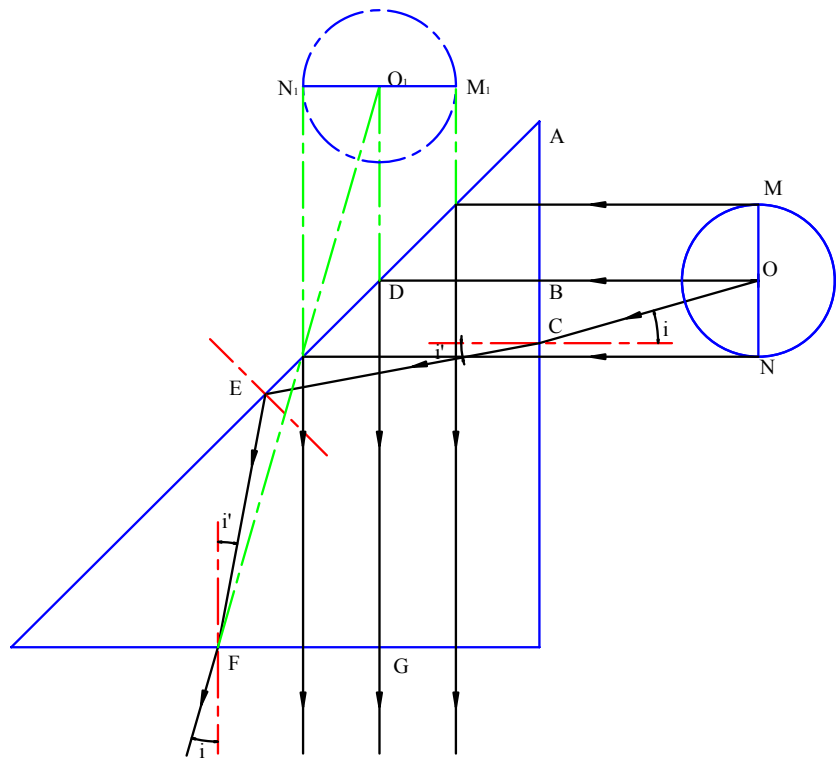

Fig. 2 Optical pathway diagram of the 3D optical system (Incident rays OB and OC are from the object point O. The two incident rays enter into the prism and exit at point F and G of the other right-angle edge of the prism. Points $O_1$, $M_1$, $N_1$ and O, M, N are the corresponding image points and object points which will be recorded synchronously by the camera)

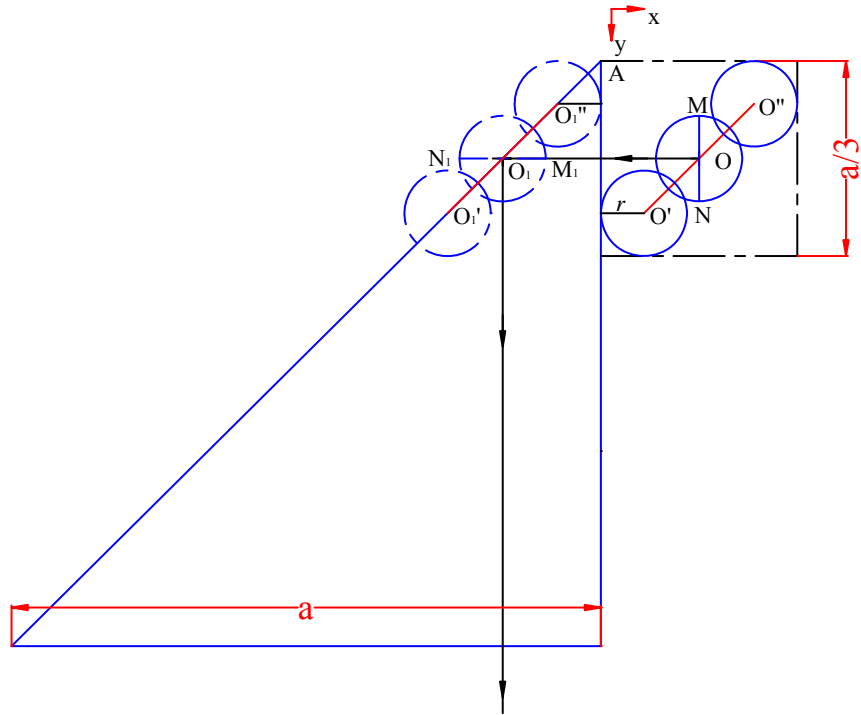

Fig. 3 The relative position of the micro-tube with respect to the prism (A is the origin in the coordinates in which y increases downward. The position of the micro-tube falls on the diagonal of the square with the side length of a/3 according to Eq. (3) in Section 2 when n=1.5 is used.)

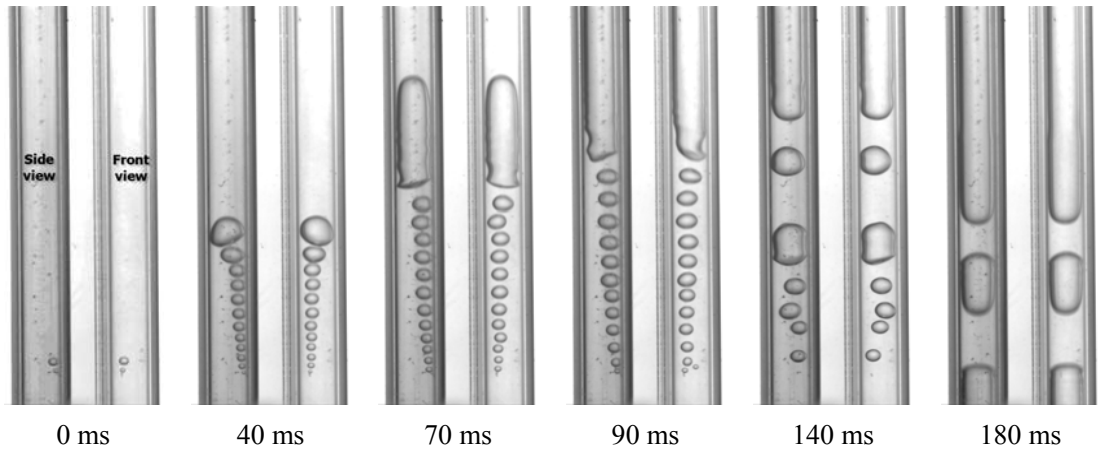

(a) Typical bubble dynamics

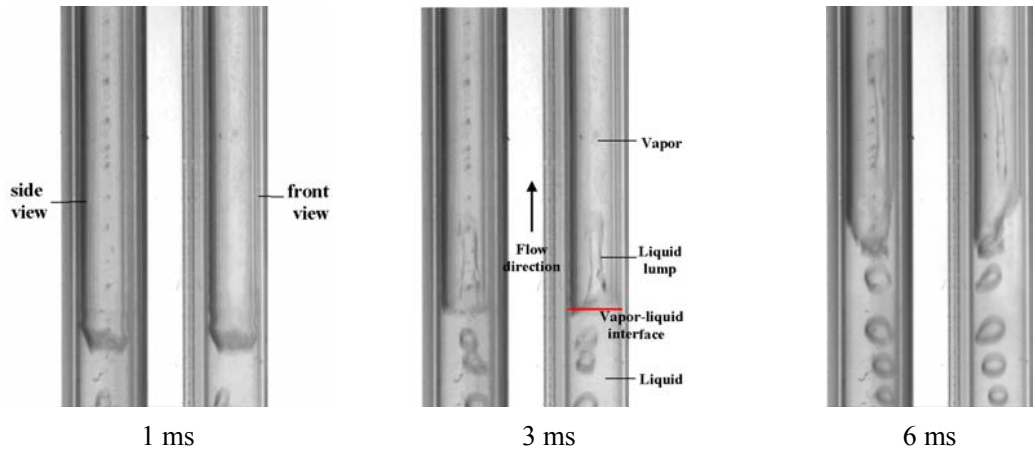

(b) liquid entrainment

Fig. 4 Typical visualization results of the two-phase flow in a micro-tube with the optical system (Liquid nitrogen, $D_{in}$=1.46mm, 1000 frames/s, the left view is the side view and right view is the front view.)

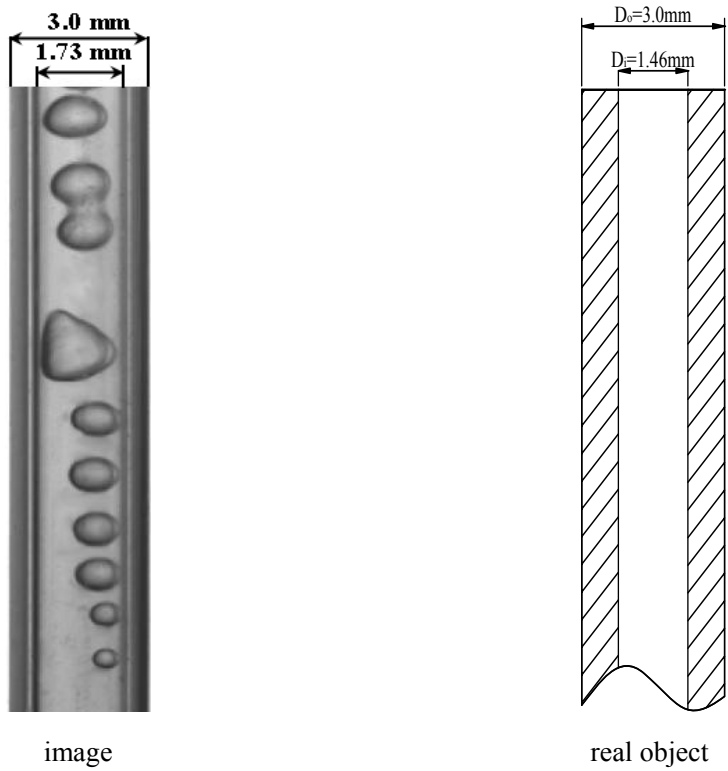

Fig. 5 Image deformation validated by the experimental result (The inner diameter of real object is 1.46 mm, whereas the inner diameter of the corresponding image increases to 1.73 mm, nearly 20% increased in size.)

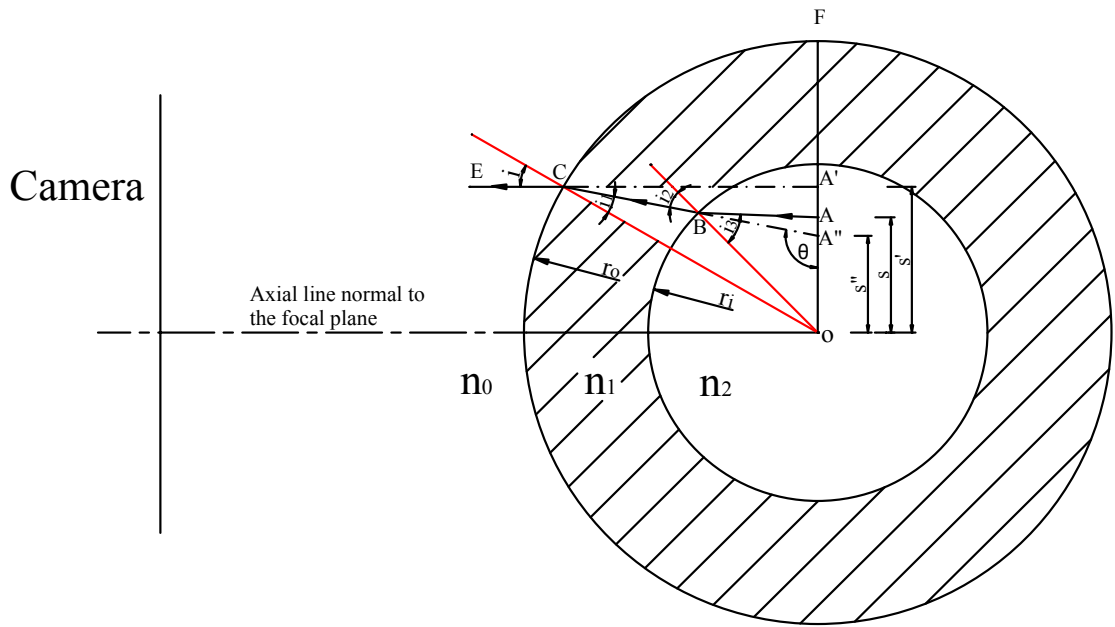

Fig. 6 Bird view of the optical path showing the image deformation (Ray AB from object point A penetrates the curved glass wall by twice refraction on the inner and outer surfaces of the glass tube, and then exits the glass tube as ray line CE in parallel direction. $s$ denotes the distance between the center O and the object point A; $s'$ denotes the distance between the center O and the image point $A'$)

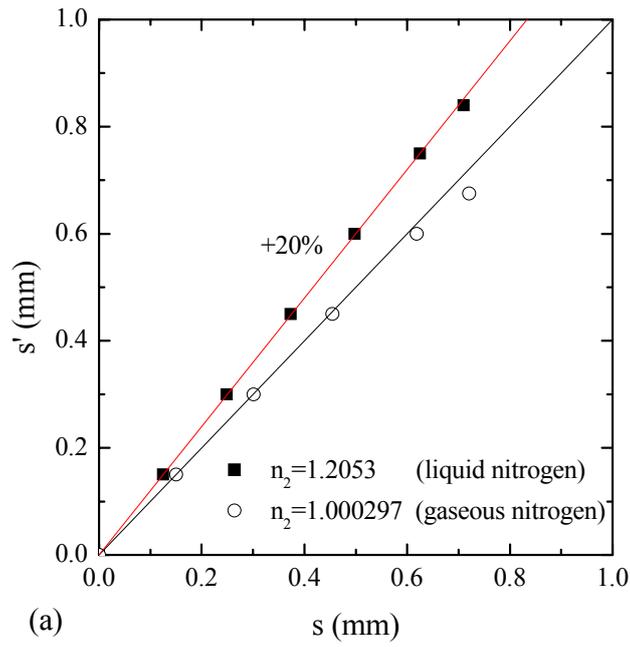

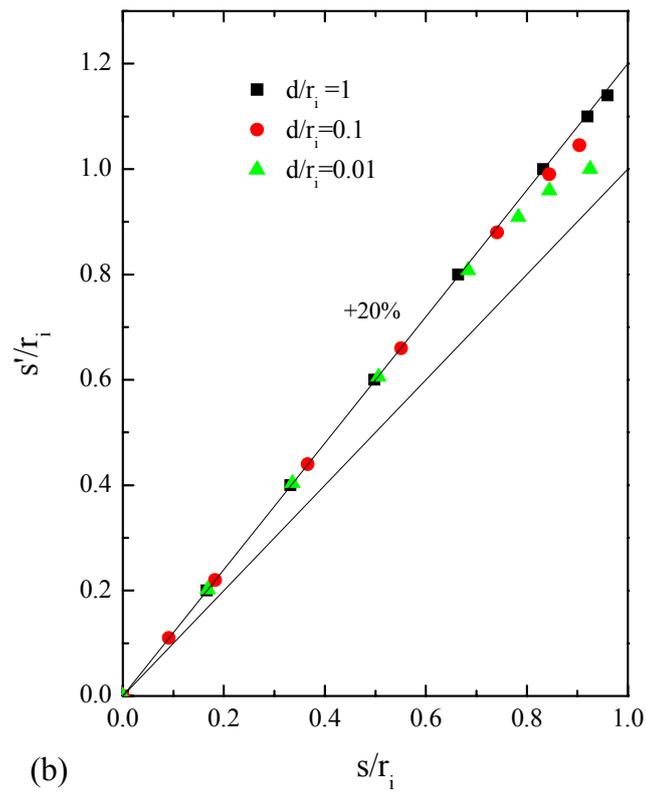

Fig. 7 Image deformation due to: (a) the effect of refraction index ($n_0$=1.0, $n_1$=1.517, $r_i$=0.73 mm, $r_o$=1.5 mm); (b) the effect of the ratio of wall thickness to the inner radius ($n_0$=1.0, $n_1$=1.517, $n_2$=1.2053)

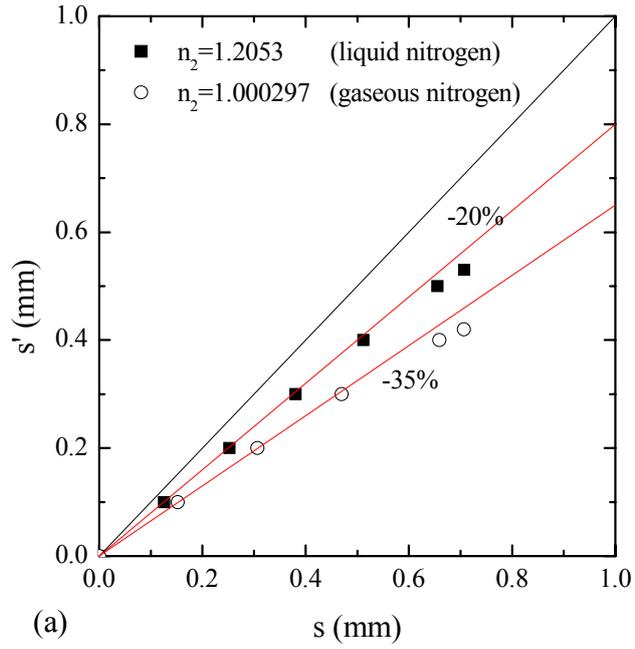

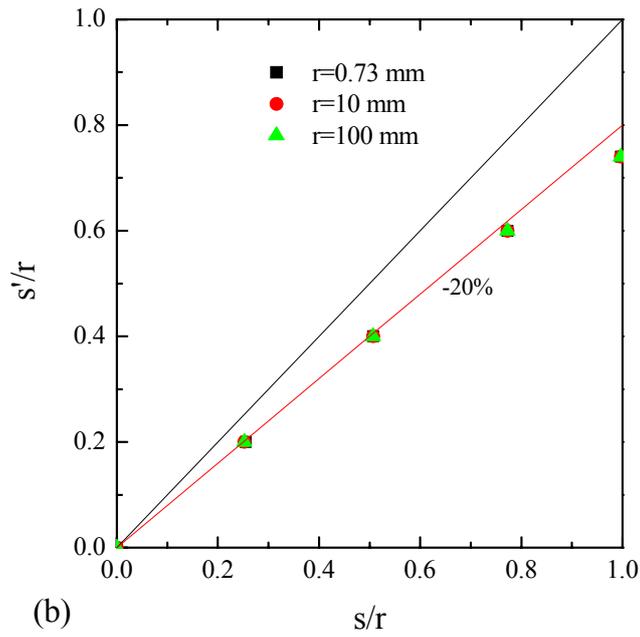

Fig. 8 Image deformation by a rectangular optical correction box: (a) the effect of refraction index ($n_0$=1.0, $n_1$=1.517, r=0.73 mm); (b) the effect of radius ($n_0$=1.0, $n_1$=1.517, $n_2$=1.2053)

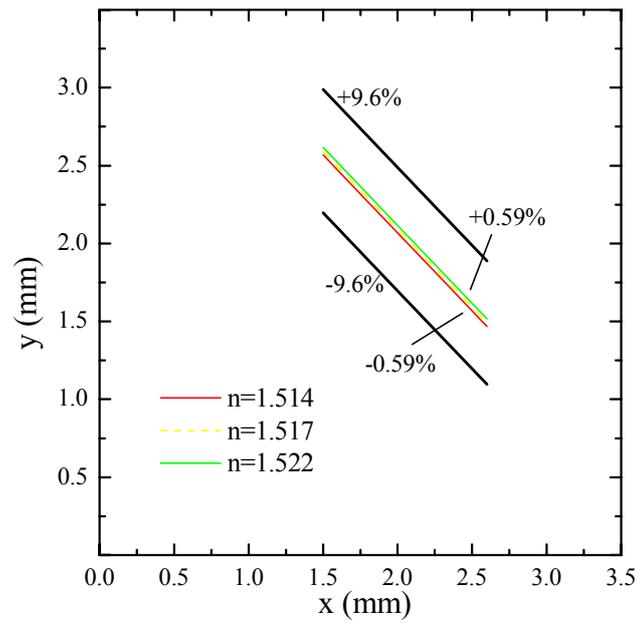

Fig. 9 Deviation due to chromatic aberration and the region where the micro-tube can be installed for clear imaging (the region for clear imaging is bordered by the two curves of ±9.6%.)

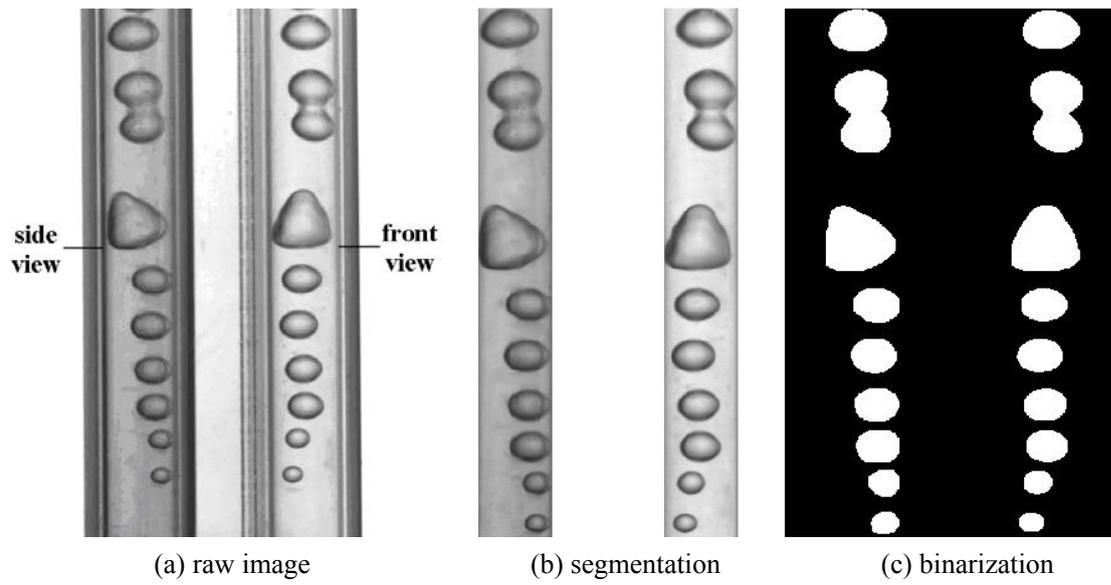

(a) raw image    (b) segmentation    (c) binarization

Fig. 10 3D re-construction step 1: the raw image is segmented firstly and then binarized by a series of image processing algorithms

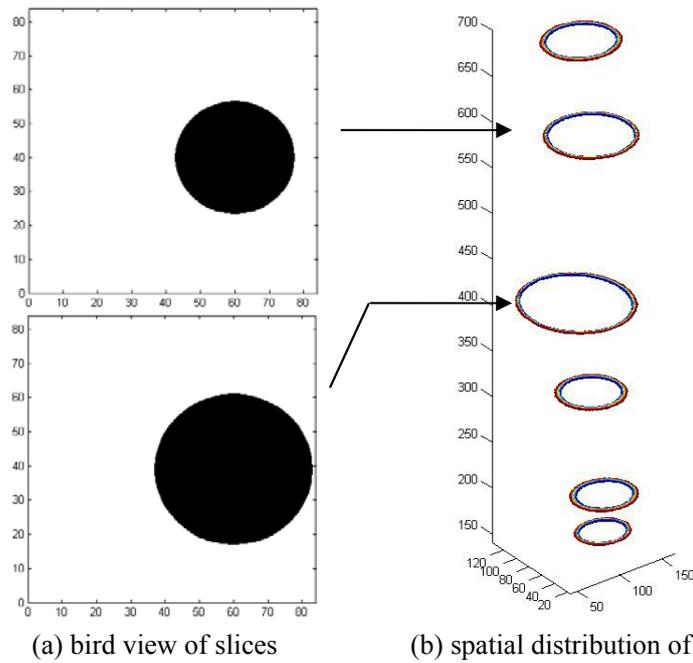

(a) bird view of slices    (b) spatial distribution of slices

Fig. 11 3D re-construction step 2: elliptical slice is drawn based on the major axis and minor axis obtained from Fig. 10(c); the two slices shown in (b) are the corresponding bird views in (a); all slices are produced in pixel coordinate system. (The units of the co-ordinates are in pixel)

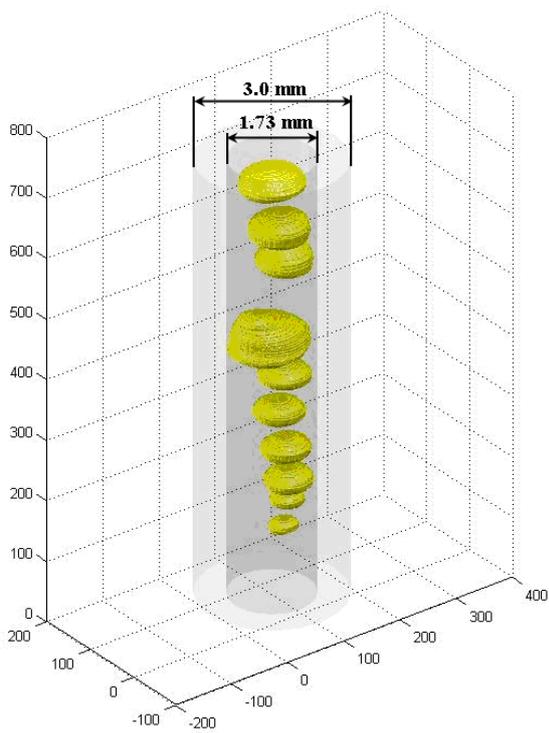 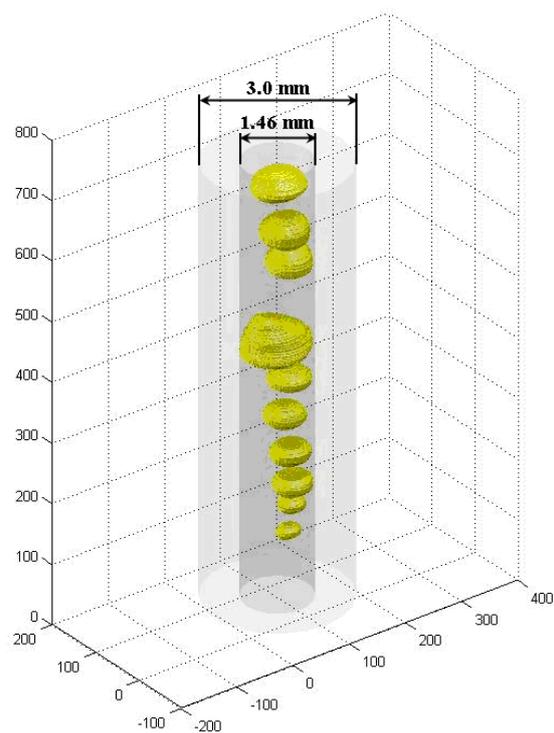

(a) 3D image	(b) corrected 3D real object

Fig. 12 3D re-construction step 3: the 3D image (left) is re-constructed by piling up slices from the Fig. 11; according to the deviation calculation in Section 3, the corresponding major axis and minor axis obtained from Fig. 10(c) can be rectified to produce new elliptical slices, which are piled up to produce the corrected 3D real object (right). (The units of the co-ordinates are in pixel)

**Appendix A**

Fig. A-1 is the optical pathway diagram of the optical system. Incident rays OB and OC are from the object point O. The two incident rays enter into the prism and reflect off the bevel edge of the prism. Finally, the rays exit at point F and G of the other right-angle edge. Points O' and $O_2$ are two image points can be observed in the prism. Points $O_1$ and O are the corresponding image point and object point which will be recorded by the camera.

Ray OB is normal to the receiving surface and the corresponding incident angle and the refractive angle are equal. For ray OC, the incident angle and the refractive angle are $i$ and $i'$, satisfying the law of refraction as below

$$\frac{\sin i}{\sin i'} = n \tag{1}$$

where $n$ is the refractive index of the prism.

The distance between point B and C is $h$ and there are the following expressions

$$\tan i = h/L, \quad \tan i' = h/L' \tag{2}$$

$L$ is the distance between point O and B, and the distance from the image point O' to the right-angle edge of the prism is $L'$, which is expressed as

$$L' = \frac{\tan i}{\tan i'} L = L \frac{n\sqrt{1 - n\sin^2 i}}{\cos i} \tag{3}$$

For paraxial rays ($i \approx 0$), there is $\cos i \approx 1$, $\sin^2 i \approx 0$. Eq. (3) can be reduced to

$$L' = nL \tag{4}$$

Both rays reflect off the bevel edge of the prim and follow the law of reflection

$$\overline{O_2 D} = \overline{O'D} = L' + y \tag{5}$$

where $y$ is the distance between points A and B. A is the origin of the coordinates, in which $y$ increases downward.

Ray OB exits in the direction of normal to another right-angle edge of the prism, the other at an angle of $i$

$$\tan i = \frac{\overline{GF}}{\overline{GO_1}}, \quad \tan i' = \frac{\overline{GF}}{\overline{GO_2}} \tag{6}$$

There is the following expression

$$\overline{GO_2} = \frac{\tan i}{\tan i'}\overline{GO_1} = \overline{GO_1}\frac{n\sqrt{1-n\sin^2 i}}{\cos i} \tag{7}$$

$$\overline{GO_2} = n \times \overline{GO_1} \tag{8}$$

The distance between exit point G and the image point $O_2$ is

$$\overline{GO_2} = \overline{GD} + \overline{O_2D} = \overline{GD} + L' + y = a + nL \tag{9}$$

where $a$ is the length of the right angle edge of the prism.

The distance between exit point G and the image point $O_1$ is

$$\overline{GO_1} = (a+nL)/n = a/n + L \tag{10}$$

The object point O and the image point $O_1$ can be clearly recorded synchronously by the camera when the two points is at the same horizontal line, i.e.

$$\overline{GO_1} = \overline{GD} \tag{11}$$

$$a/n + L = a - y \tag{12}$$

Then the following formula, which governs the location of the axis of the micro-tube, can be written as

$$y = a(n-1)/n - x \tag{13}$$

**Appendix B**

Shown in Fig. 6 is the bird view of the optical path showing image deformation. The angle of incidence

$i$ is given by

$$i = \arcsin(s'/r_o) \quad (1)$$

For ray EC, the incident angle and the refractive angle are $i$ and $i_1$, satisfying the law of refraction as below

$$\sin i \times n_0 = \sin i_1 \times n_1 \quad (2)$$

$$i_1 = \arcsin((s'/r_o) \times n_0 / n_1) \quad (3)$$

where $n_0$ and $n_1$ are the refractive indexes of air and the glass, respectively.

Prolong the lines EC and CB east to intersect the line OF at the points $A'$ and $A''$. Then apply the law of sines in the triangle $OCA''$,

$$\frac{s''}{\sin i_1} = \frac{r_o}{\sin \theta} \quad (4)$$

where

$$\theta = \pi/2 - i_1 + i \quad (5)$$

$s''$ is given by

$$s'' = \frac{r_o \sin i_1}{\cos(i_1 - i)} \quad (6)$$

Ray CB is refracted at the inner surface of the tube,

$$\sin i_2 \times n_1 = \sin i_3 \times n_2 \quad (7)$$

Applying the law of sines to the triangle $OBA''$

$$\frac{s''}{\sin i_2} = \frac{r_i}{\sin \theta} \quad (8)$$

Obtaining

$$i_2 = \arcsin(\frac{s''\sin\theta}{r_i}) \qquad (9)$$

$$i_3 = \arcsin(\frac{n_1 \sin i_2}{n_2}) \qquad (10)$$

$s$ is obtained by applying the law of sines to the triangle $OBA$

$$s = \frac{r_i \sin i_3}{\sin(\theta + i_2 - i_3)} \qquad (11)$$

Fig. B-1 shows the optical path of the method with rectangular optical correction box. Parallel ray EB enters into the glass box and experiences refraction at the inner curve wall, finally intersects the line OF at point A. Point A is the object point.

The angle of incidence $i$

$$i = \arcsin(s'/r) \qquad (12)$$

For ray EB, the incident angle and the refractive angle are $i$ and $i_1$, satisfying the law of refraction as below

$$\sin i \times n_1 = \sin i_1 \times n_2 \qquad (13)$$

$$i_1 = \arcsin((s'/r) \times n_1 / n_2) \qquad (14)$$

Prolong the line EB east to intersect the line OF at the point $A'$, which is the image point. Then apply the law of sines to the triangle $OAB$

$$\frac{s}{\sin i_1} = \frac{r}{\sin\theta} \qquad (15)$$

$$s = \frac{r \sin i_1}{\sin\theta} \qquad (16)$$

where

$$\theta = \pi/2 - i_1 + i \tag{17}$$

**Appendix C**

Abbe number ($V$) is adapted to assess the chromatic aberration,

$$V = \frac{n_d - 1}{n_F - n_C} \tag{1}$$

where $n_d$, standard refractive index which corresponds to yellow light to which the naked eye is most sensitive; $n_F$: the refraction index of blue light; $n_C$: the refraction index of red light;

Cauchy equation:

$$n(\lambda) = A + \frac{B}{\lambda^2} + \frac{C}{\lambda^4} \tag{2}$$

where

$$A = n_d - \frac{B}{\lambda_d^2}, \quad B = \frac{(n_d - 1)\lambda_C^2 \lambda_F^2}{V(\lambda_C^2 - \lambda_F^2)} \tag{3}$$

$\lambda_C$: wavelength of red light; $\lambda_F$: wavelength of blue light; $\lambda_d$: wavelength of yellow light.

According to Schott® Glass database, which offers Abbe number and standard refractive index for various types of glass, the parameters for the prism of the present study are: Type: BK7, $n_d$=1.51680, $V$ = 64.17.

Other parameters are calculated by neglecting the third term on the right side of Eq. (2) as follows:

$$A = 1.5168 - \frac{4.2173773 \times 10^3}{587.5618^2} = 1.5045838 \tag{4}$$

$$B = \frac{(1.51680 - 1) \times 656.2725^2 \times 486.1327^2}{64.17 \times (656.2725^2 - 486.1327^2)} = 4.2173773 \times 10^3 \text{ nm}^2 \tag{5}$$

$$n_C = 1.5045838 + \frac{4.2173773 \times 10^3}{656.2725^2} = 1.514 \tag{6}$$

$$n_F = 1.5045838 + \frac{4.2173773 \times 10^3}{486.1327^2} = 1.522 \tag{7}$$

Substituting the calculated refractive indexes ($n_C, n_d, n_F$) to Eq. (13) in appendix A results in three equations. The corresponding three curves for the equations are shown in Fig. 9, where the length of the right angle edge of the prism is 12 mm. Maximal deviation due to chromatic aberration is estimated as:

$$\overline{GO_1}(n_C) - \overline{GO_1}(n_F) = a(1/n_C - 1/n_F) = 0.048 \text{mm} \tag{8}$$

where $\overline{GO_1}$ is according to Eq. (10) in Appendix A.

For the chosen macro lens in the present optical system, depth of field is 0.396 mm for the case of f/2.8 aperture and magnification of 1x. The calculated deviation distance of 0.048 mm is within the range of depth of field (0.396 mm) in the present optical system. Therefore, the unsharpness of the image caused by chromatic aberration is imperceptible and the white light can be selected as the light source in the present study.

The uncertainty of the region where the micro-tube can be installed for clear imaging is estimated based on depth of field (DOF). Although a lens can precisely focus at only one distance, the decrease in sharpness is gradual on either side of the focused distance, so that within the depth of field, the unsharpness is imperceptible. According to depth of field of the adopted lens in the study, the region, where the unsharpness of the image is imperceptible, is obtained by Eq. (13) in appendix A:

$$y = a(n-1)/n - x \pm k \tag{9}$$

Where $k$ is the depth of field of the lens in the present study. The corresponding two curves are plotted in Fig. 9 in the case of $k$ equal to 0.396 mm. It shows that clear image can be obtained in the region bordered by the two curves of $\pm 9.6\%$.

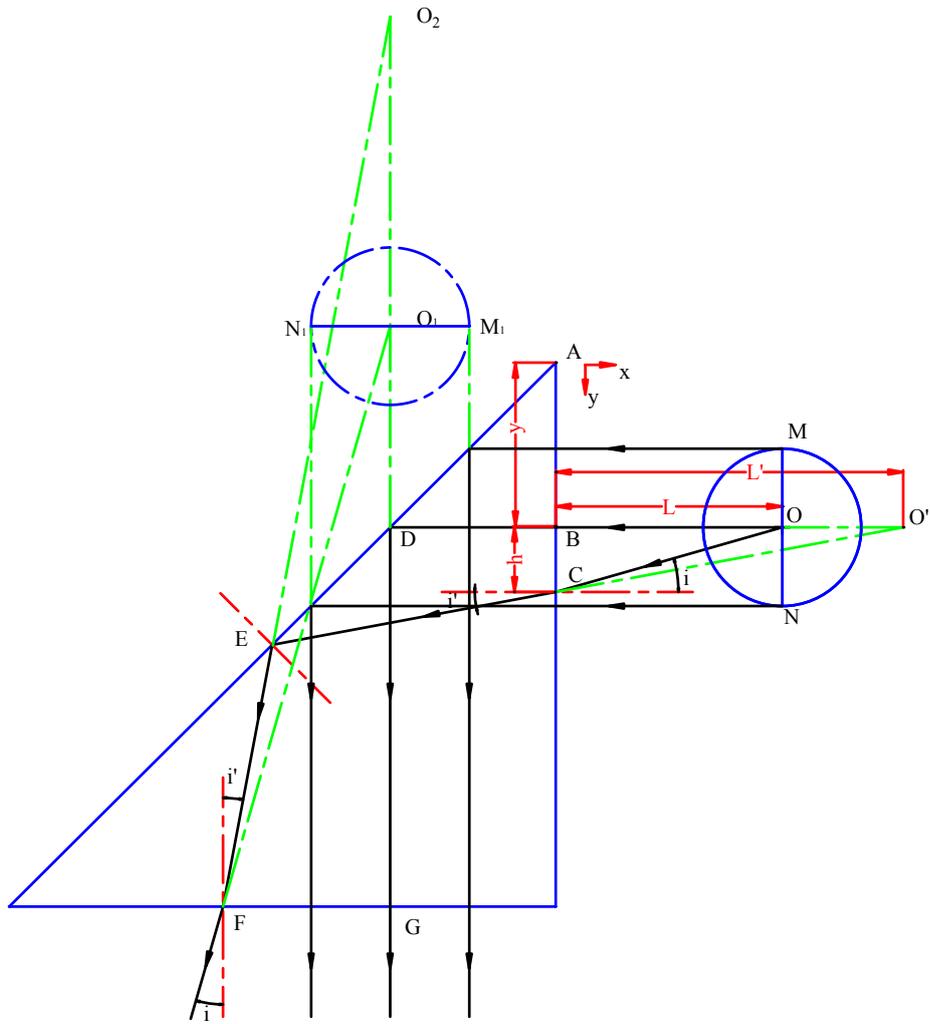

Fig. A-1 Detailed optical pathway diagram of the prism system

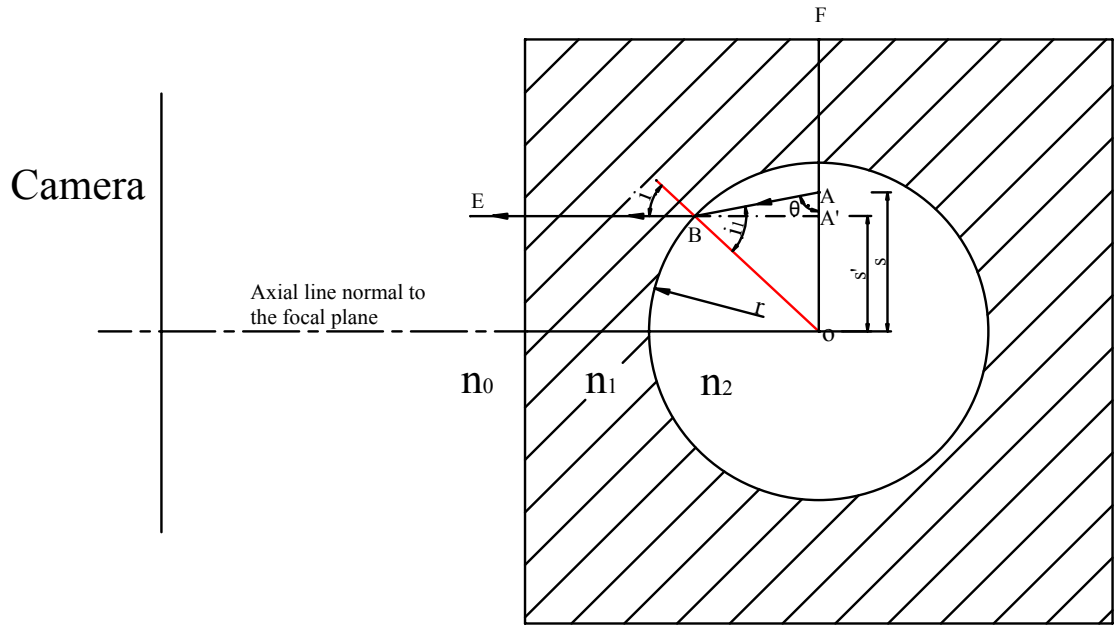

Fig. B-1 Bird view of the optical path showing the image deformation with a rectangular optical correction box